\documentclass[runningheads]{llncs}
\usepackage[T1]{fontenc}
\usepackage{graphicx}
\usepackage[numbers]{natbib} 
\usepackage{amssymb}
\usepackage{amsmath}
\usepackage{graphicx}
\usepackage{algorithm}
\usepackage{algpseudocode}

\begin{document}
\title{Towards Robust Speech Recognition for Jamaican Patois Music Transcription}
%
%
\author{Jordan Madden \and
Matthew Stone\and
Dimitri Johnson
\and
Daniel Geddez}
\authorrunning{J. Madden et al.}
%
\institute{Jamaica Artificial Intelligence Association \newline Kingston, Jamaica \newline
\email{admin@jaia.org.jm}\\}
\maketitle              
\begin{abstract}
Although Jamaican Patois is a widely spoken language, current speech recognition systems perform poorly on Patois music, producing inaccurate captions that limit accessibility and hinder downstream applications. In this work, we take a data-centric approach to this problem by curating more than 40 hours of manually transcribed Patois music. We use this dataset to fine-tune state-of-the-art automatic speech recognition (ASR) models, and use the results to develop scaling laws for the performance of Whisper models on Jamaican Patois audio. We hope that this work will have a positive impact on the accessibility of Jamaican Patois music and the future of Jamaican Patois language modeling.

\end{abstract}
\section{Introduction}

Jamaican Patois, also known as Jamaican Creole, is a primarily spoken language with widespread use across Jamaica and the global Caribbean diaspora. Despite its global reach and prevalence in music, and everyday communication, Patois remains severely underrepresented in modern language technologies. In particular, automatic speech recognition (ASR) systems perform poorly when applied to Patois speech. This is largely because Patois is a low-resource language, with limited transcribed speech data available.

One area where this limitation is especially apparent is in music transcription. Music is a central vehicle for Jamaican cultural expression, with many songs being performed entirely in Patois. However, existing ASR systems, such as the automatic captions provided by YouTube or other media platforms, fail to accurately transcribe the Patos audio. The resulting captions are often inaccurate, and in some cases completely unrelated to the actual lyrics being sung. This hinders accessibility for people who are not fluent in Patois, including members of the deaf and hard-of-hearing community, as well as international audiences interested in Jamaican music and culture.

Beyond accessibility, the inability to transcribe Jamaican Patois also poses a significant barrier to the development of downstream natural language processing (NLP) technologies. Large language models (LLMs), which have transformed the field of AI, rely heavily on large-scale datasets that are high-quality and diverse. For languages like Patois, which are predominantly spoken, creating such datasets is nearly impossible without robust speech-to-text systems. A high-quality Patois ASR model could therefore be important in enabling the construction of foundational LLMs for Jamaican Patois.

In this work, we take steps toward building robust ASR systems for Jamaican Patois in musical contexts. Our contributions are threefold: we introduce a supervised dataset comprising approximately 42 hours of transcribed Patois music (the largest of its kind as far as we are aware), we fine-tune a series of Whisper models on our dataset to evaluate ASR performance, and we develop a scaling equation that model ASR performance as a function of dataset size and model capacity.  By improving the accuracy and reliability of Patois transcription, we hope to contribute to the broader ecosystem of tools for making underrepresented languages more accessible and better supported in AI systems.

\section{Related Work}

\subsection{Low-Resource Automatic Speech Recognition (ASR)}

Automatic Speech Recognition (ASR) has undergone significant advances with the introduction of large-scale models such as Whisper~\cite{radford2023robust}, wav2vec 2.0~\cite{baevski2020wav2vec}, and HuBERT~\cite{hsu2021hubert}, which are pre-trained on massive multilingual datasets. However, these models often underperform on dialects and languages with limited labeled data due to a mismatch in phonetic and linguistic features.

To address this, several approaches have been proposed. Transfer learning has been effective for low-resource ASR by fine-tuning pre-trained models on smaller domain-specific datasets~\cite{karunanayake2019transfer}. Domain adaptation methods~\cite{anoop2021unsupervised} attempt to bridge the distribution gap between high- and low-resource domains. Data augmentation techniques~\cite{ko2015audio}, such as speed perturbation and noise injection, can help to improve robustness in limited-data settings.

\subsection{Language Models for ASR Post-processing}

Large Language Models (LLMs) like GPT4 have been used to improve ASR outputs by post-processing or correction. Recent studies~\cite{lefevre2025llm} have shown that prompting LLMs to correct noisy transcripts can reduce Word Error Rate (WER), especially in low-resource or domain-specific audio such as music or medical speech. 

\subsection{Music Transcription and Code-Switching ASR}

ASR for music transcription poses unique challenges including background instrumentation, rhythm, and non-standard pronunciation. Prior work has focused on English-language lyric transcription~\cite{gao2022automatic}, but there has been limited progress on music in creole or code-switched languages. Work like SUPERB~\cite{yang2021superb} aims to evaluate ASR performance when trained on vast amounts of data in a similar way as LLMs but still lack support for dialectal music.

In multilingual and code-switching scenarios, ASR models struggle due to overlapping phonetic inventories and frequent language shifts. Efforts like~\cite{biswas2022code} highlight the need for datasets and models specifically tuned to these dynamics. Jamaican Patois, with its unique linguistic structure and English influences, presents a hybrid challenge of dialectal ASR and code-switching.

\subsection{Scaling Laws in ASR and NLP}

Scaling laws, which model performance as a function of model size and dataset size, have gained traction in language modeling~\cite{kaplan2020scaling} and are beginning to appear in speech~\cite{droppo2021scaling}. These laws provide guidance on resource allocation and model design, especially when data is scarce. However, their application to dialectal and music-based ASR remains unexplored. Our work contributes to this space by empirically deriving scaling laws for Whisper models fine-tuned on Jamaican Patois music, enabling better planning for model deployment in low-resource settings.

\subsection{LLMs for Music Content Filtering and Restoration}

Stone et al.~\cite{stone2024ai} explored the use of ASR systems and large language models (LLMs) to automatically clean and filter explicit lyrics in Jamaican music. Their system leveraged an ASR model to convert the audio to text and then used LLM to detect lewd content in transcriptions of music and propose sanitized alternatives.

Our work is directly related to their efforts as the authors proposed the use of the Whisper models to perform the transcription from audio to text. Since their focus was specifically on Jamaican music, the models that we train in this work will be able to improve the performance of their proposed system as we later show that our models outperform standard Whisper models on Jamaican Patois.

\section{Approach}

\subsection{Problem Formulation}

Consider a collection of real-valued audio clips $\{a_1, a_2, ..., a_n\}$ and their corresponding text transcriptions tokens $\{t_1, t_2, ..., t_n\}$,   our goal is to learn a function $F: \mathbb{R}^n \rightarrow  \mathbb{I}^n$ such that $x_i = F(a_i)$ and that $||t_i - x_i|| \le \epsilon$ where $\epsilon$ is small. 

\subsection{Dataset}
The dataset consists of 5,110 recordings of Jamaican music accompanied by corresponding Jamaican Patois transcriptions. Each data point includes a URL linking to a 30-second MP3 audio clip, a manually annotated transcription of the audio segment, and the official lyrics of the full song. The audio is sampled at 22,050 Hz, resulting in 661,500 samples per recording. In total, the dataset comprises 42.58 hours of audio. To facilitate ease of use with automatic speech recognition (ASR) systems, we develop a data processing script that converts the raw data into a format suitable for ingestion by popular deep learning frameworks such as PyTorch \cite{paszke2019pytorch} and HuggingFace Transformers \cite{wolf2020transformers}. This dataset was initially annotated by hand and is intended to support research in the captioning and transcription of Jamaican music. To the best of our knowledge, it represents the largest publicly available dataset of its kind.

\subsection{Model Finetuning}
We utilize OpenAI’s Whisper models, accessed via the Huggingface Transformers library, as the foundation for our transcription pipeline. The Whisper family of models is pre-trained on 680,000 hours of multilingual audio, enabling it to learn robust and transferable audio representations. We hypothesize that these features will facilitate effective adaptation to Jamaican Patois through fine-tuning. To this end, we fine-tune the ‘tiny’, 'base',  ‘small’, and ‘medium’ variants of Whisper using our dataset. We did not fine-tune the 'large' variants of the Whisper models due to computational constraints. Details about how the models were trained and the results of the experiments can be found in Section \ref{results}.

\section{Results}\label{results}

\begin{figure}[t]
\includegraphics[width=\textwidth]{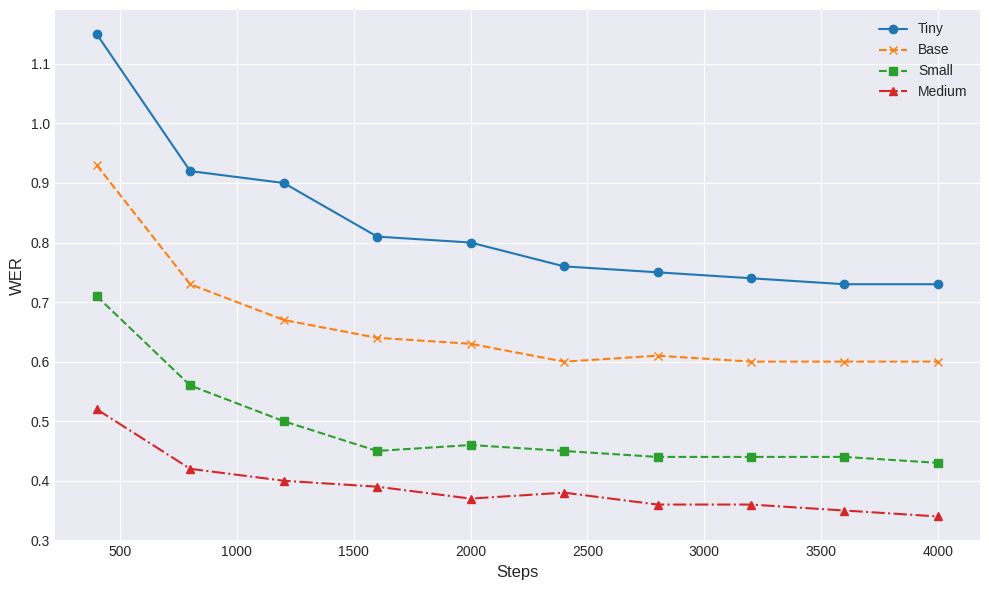}
\caption{Word Error Rate vs Training Steps for various sizes of the Whisper Models.} \label{wer_size}
\end{figure}
\subsection{Experimental Settings}

We fine-tuned the various Whisper models using various amounts of the aforementioned dataset. For each Whisper variant that we trained, we trained on 20, 35, and 40 hours of audio respectively.  The models are optimized using the AdamW optimizer, with an initial learning rate of 0.00001. We varied this throughout training using a Linear Learning Rate Scheduler with a warmup phase of 500 steps. Training is conducted for 4,000 steps in total, with the learning rate warming up over the first 500 steps. Prior to fine-tuning, all audio is resampled from 22,050 Hz to 16,000 Hz to match Whisper’s expected input format and then transformed into log Mel-spectrograms. The log Mel-spectrograms are passed as input to the Whisper models.

To evaluate the quality of the model, we use of the Word Error Rate metric. It is defined as follows:
\begin{equation}
    WER(x, y) = \frac{1}{N}\sum_{i=1}^N
    \mathbb{I}(x_i, y_i)
\end{equation}
where $N$ is the number of words in a sample transcription, $i$ represents the index of a word in the transcription, $x$ is the generated transcript, $y$ is the ground truth transcript, and $\mathbb{I}(\cdot)$ is the indicator function  $\mathbb{I}(x, y)=1$ if $x\neq y$ and 0 if $x= y$. Additionally, to measure the loss during training and validation, we made use of the Cross-Entropy Loss function \cite{shannon1948mathematical}.
    
\subsection{Fine-Tuning Results}

    Here, we fine-tuned some of the Whisper family of models in our data set. Figure \ref{wer_size} shows how the Word Error Rate (WER) changes as training progresses for all model sizes (Tiny, Base, Small, and Medium) when trained on 35 hours of audio. As expected, all models get better (i.e., WER goes down) the more they train, and the bigger models clearly perform better overall. This is in keeping with the results presented in \cite{kaplan2020scaling} that show that larger models exhibit more sample efficiency when learning than smaller models. The Medium model starts with a WER around 0.52 and steadily drops to about 0.34 by 4000 steps. The Small model also improves, ending up around 0.45, and the Base model ends up around 0.60. The Tiny model, on the other hand, starts off quite high (above 1.1) and ends up leveling off near 0.73. While all four models benefit from training, the Medium and Small models learn faster and more efficiently, with their WERs stabilizing earlier, this is an expected result because these are the 2 largest of the models tested.. This idea is further underscored by the results in Table \ref{tab:wer_data}.

We also compared the performance of the fine-tuned Whisper models to the pre-trained Whisper Large model. This model was not finetuned as the other models in the Whisper family were. On standard English, Whisper Large (the most performant model in the Whisper family) has a WER of approximately 0.05, while on Jamaican Patois, it has a WER of 0.89 as seen in Table \ref{tab:wer_data}. This goes to show that even though Jamaican Patois is related to English, the priors provided by Whisper are not sufficient to effectively perform transcription on Jamaican Patois. From our results, we can see that even the Whisper Tiny model, which is approximately 50x smaller than the Whisper Large model, can outperform it by a significant margin. The scaling results that are demonstrated in our experiments make us confident that further increases in model size/parameter count will elicit better WER performance.

\begin{table}[t]
    \centering
        \caption{Best Word Error Rate obtained. Tiny, Base, Small, and Medium models were fine-tuned while Large was not. Large is included here as a benchmark.}
    \begin{tabular}{|c||c|c|c|c|}\hline
 & \multicolumn{4}{|c|} {$\textbf{Best WER}$ $\downarrow$} \\\hline
        $\textbf{Hours}$& Whisper Tiny& Whisper Base & Whisper Small & Whisper Medium  \\ \hline\hline
        20 & 0.79 & 0.69 & 0.51 & 0.40  \\ \hline
        35 & 0.73 & 0.60 & 0.44 & 0.34  \\ \hline
        40 & 0.70 & 0.58 & 0.41 & 0.30  \\ \hline\hline
 Whisper Large*& \multicolumn{4}{|c|}{0.89}\\\hline
    \end{tabular}

    \label{tab:wer_data}
\end{table}

\subsection{Scaling Characteristics}

\begin{figure}[t]
\includegraphics[width=\textwidth]{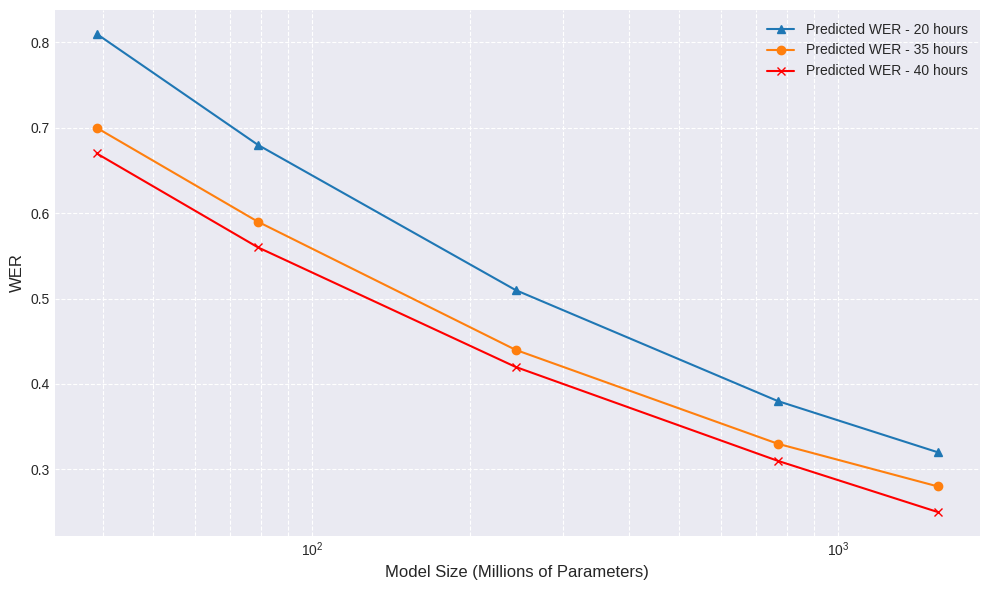}
\caption{Predicted Word Error Rate vs Model Size on a Logarithmic Scale} \label{wer_size}
\end{figure}
Building on the results of the previous section, we aim to develop a principled way to predict how well a model will perform given its size and the amount of training data available. To this end, we analyze the scaling characteristics of Whisper models fine-tuned on Jamaican Patois music and try to quantify their performance.

We hypothesize that the WER scales as a power-law function of both the model size $M$ (measured in number of parameters) and the dataset size $D$ (measured in hours), and we express this relationship as:

\begin{equation}
    \text{WER} = A \cdot M^{-\alpha} \cdot D^{-\beta}
\end{equation}

Taking the natural logarithm of both sides yields a linear relationship:

\begin{equation}
    \log(\text{WER}) = \log(A) - \alpha \cdot \log(M) - \beta \cdot \log(D)
\end{equation}

Here:
\begin{itemize}
    \item $\alpha$ captures how sensitive WER is to increases in model size,
    \item $\beta$ captures how sensitive WER is to increases in dataset size,
    \item and $A$ is a scaling constant.
\end{itemize}

Using this log-log formulation, we perform linear regression to fit the parameters $A$, $\alpha$, and $\beta$ to our experimental results. The best-fitting equation we obtain is:

\begin{equation}
    \log(\text{WER}) = 5.063 - 0.255 \cdot \log(M) - 0.269 \cdot \log(D)
\end{equation}
\begin{equation}
    \text{WER} = 158.06 \cdot M^{-0.255} \cdot D^{-0.269}
\end{equation}

For each combination of dataset size and Whisper model variant used in our experiments, we went on to calculated the predicted WER using the derived scaling law. These predictions are visualized in Figure \ref{wer_size}. As shown, the predicted WER values closely align with the empirical results reported in Table \ref{tab:wer_data}, providing strong validation for the accuracy of our scaling model. This agreement also lends credibility to our estimate for the performance of the Whisper Large model, despite it not being fine-tuned in our experiments. Beyond validation, this scaling law can serve as a practical tool for guiding future decisions about model selection and dataset size given the computational resources available to someone and their desired performance.

\section{Conclusion}
We addressed the lack of accessible, high-quality transcriptions for Jamaican Patois music by curating a new dataset of over 40 hours of annotated music and proposing a scalable data flywheel to generate additional training data with minimal human input. We also fine-tuned Whisper models to demonstrate the utility of the dataset and found that there was a clear inverse relationship between the model size/parameter count and the WER. Building on those observations, we developed scaling laws that can predict the performance of a Whisper model of a given size that is trained on a given amount of data. We hope that this work moves us closer to building reliable ASR systems for Jamaican Patois, improving accessibility for Patois audio content, and laying the foundations for improved Jamaican Patois language modeling. 

\begin{credits}

\subsubsection{\discintname}
The authors have no competing interests to declare that are
relevant to the content of this article.
\end{credits}

 \bibliographystyle{splncs04}
 \bibliography{references}

\begin{thebibliography}{10}
\providecommand{\url}[1]{\texttt{#1}}
\providecommand{\urlprefix}{URL }
\providecommand{\doi}[1]{https://doi.org/#1}

\bibitem{anoop2021unsupervised}
Anoop, C.S., Prathosh, A., Ramakrishnan, A.: Unsupervised domain adaptation schemes for building asr in low-resource languages. In: 2021 IEEE Automatic Speech Recognition and Understanding Workshop (ASRU). pp. 342--349. IEEE (2021)

\bibitem{baevski2020wav2vec}
Baevski, A., Zhou, Y., Mohamed, A., Auli, M.: wav2vec 2.0: A framework for self-supervised learning of speech representations. Advances in neural information processing systems  \textbf{33},  12449--12460 (2020)

\bibitem{biswas2022code}
Biswas, A., Y{\i}lmaz, E., van~der Westhuizen, E., de~Wet, F., Niesler, T.: Code-switched automatic speech recognition in five south african languages. Computer Speech \& Language  \textbf{71},  101262 (2022)

\bibitem{droppo2021scaling}
Droppo, J., Elibol, O.: Scaling laws for acoustic models. arXiv preprint arXiv:2106.09488  (2021)

\bibitem{gao2022automatic}
Gao, X., Gupta, C., Li, H.: Automatic lyrics transcription of polyphonic music with lyrics-chord multi-task learning. IEEE/ACM Transactions on Audio, Speech, and Language Processing  \textbf{30},  2280--2294 (2022)

\bibitem{hsu2021hubert}
Hsu, W.N., Bolte, B., Tsai, Y.H.H., Lakhotia, K., Salakhutdinov, R., Mohamed, A.: Hubert: Self-supervised speech representation learning by masked prediction of hidden units. IEEE/ACM transactions on audio, speech, and language processing  \textbf{29},  3451--3460 (2021)

\bibitem{kaplan2020scaling}
Kaplan, J., et~al.: Scaling laws for neural language models. arXiv preprint arXiv:2001.08361  (2020)

\bibitem{karunanayake2019transfer}
Karunanayake, Y., Thayasivam, U., Ranathunga, S.: Transfer learning based free-form speech command classification for low-resource languages. In: Proceedings of the 57th Annual Meeting of the Association for Computational Linguistics: Student Research Workshop. pp. 288--294 (2019)

\bibitem{ko2015audio}
Ko, T., Peddinti, V., Povey, D., Khudanpur, S.: Audio augmentation for speech recognition. In: Interspeech. vol.~2015, p.~3586 (2015)

\bibitem{lefevre2025llm}
LeFevre, G., Hosier, J., Zhou, Y., Gurbani, V.K.: Llm selection: Improving asr transcript quality via zero-shot prompting. In: SoutheastCon 2025. pp. 1440--1445. IEEE (2025)

\bibitem{paszke2019pytorch}
Paszke, A.: Pytorch: An imperative style, high-performance deep learning library. arXiv preprint arXiv:1912.01703  (2019)

\bibitem{radford2023robust}
Radford, A., Kim, J.W., Xu, T., Brockman, G., McLeavey, C., Sutskever, I.: Robust speech recognition via large-scale weak supervision. In: International conference on machine learning. pp. 28492--28518. PMLR (2023)

\bibitem{shannon1948mathematical}
Shannon, C.E.: A mathematical theory of communication. The Bell system technical journal  \textbf{27}(3),  379--423 (1948)

\bibitem{stone2024ai}
Stone, M., Mansingh, G.: Ai tool for cleaning up lewd music: A jamaican perspective (research-in-progress)  (2024)

\bibitem{wolf2020transformers}
Wolf, T., Debut, L., Sanh, V., Chaumond, J., Delangue, C., Moi, A., Cistac, P., Rault, T., Louf, R., Funtowicz, M., et~al.: Transformers: State-of-the-art natural language processing. In: Proceedings of the 2020 conference on empirical methods in natural language processing: system demonstrations. pp. 38--45 (2020)

\bibitem{yang2021superb}
Yang, S.w., Chi, P.H., Chuang, Y.S., Lai, C.I.J., Lakhotia, K., Lin, Y.Y., Liu, A.T., Shi, J., Chang, X., Lin, G.T., et~al.: Superb: Speech processing universal performance benchmark. arXiv preprint arXiv:2105.01051  (2021)

\end{thebibliography}

\end{document}